\documentclass[english,twocolumn, prl, noeprint, aps, superscriptaddress, numerical, footinbib]{revtex4-1}

\usepackage{amsmath}
\usepackage{babel}
\usepackage{graphicx}
\usepackage{microtype}
\usepackage{hyperref}
\hypersetup{
	colorlinks = true,
	linkcolor = blue,
	urlcolor = blue,
	citecolor = blue
}

\usepackage[utf8]{inputenc}
\usepackage{braket}

\usepackage{geometry}
\usepackage{siunitx}

\usepackage{graphicx,color}

\definecolor{mygreen}{rgb}{0,0.5,0}
\definecolor{myblue}{rgb}{0,0,0.75}
\definecolor{mymagenta}{cmyk}{0,1,0,0.12}

\newcommand{\Lop}{\hat{\mathbf{L}}}
\newcommand{\Lz}{\hat{L}_{z}}
\newcommand{\Lmin}{\hat{L}_{-}}
\newcommand{\Lpl}{\hat{L}_{+}}

\newcommand{\fParticle}{N_\text{p}/N}
\newcommand{\fGauge}{L_z/L}

\newcommand{\nNup}{N_{\uparrow}}
\newcommand{\nNdown}{N_{\downarrow}}

\newcommand{\chiSS}{\chi_s}
\newcommand{\chiNL}{\chi_{NL}}
\newcommand{\chiLL}{\chi_{L}}
\newcommand{\chiNN}{\chi_{N}}

\newcommand{\deltaS}{\delta_{ext}^s}
\newcommand{\deltaL}{\delta_{ext}^L}
\newcommand{\deltaN}{\delta_{ext}^N}

\newcommand{\deltaSS}{\delta_{int}^s}
\newcommand{\deltaLL}{\delta_{int}^L}
\newcommand{\deltaNN}{\delta_{int}^N}

\newcommand{\XOneOneS}{X^s_{11}}
\newcommand{\XOneZeroS}{X^s_{10}}
\newcommand{\XZeroZeroS}{X^s_{00}}

\newcommand{\XSCC}{X^{SCC}}

\newcommand{\bMFone}{\hat{b}_{\text{v}}}
\newcommand{\bMFzero}{\hat{b}_{\text{p}}}

\newcommand{\aBohr}{a_B}

\newcommand{\gSCC}{g^{SCC}}

\newcommand{\NaOne}{\psi_{N,1}}

\newcommand{\dNaZero}{\psi_{N,0}^\dag}

\newcommand{\dLiOne}{\psi_{L,1}^\dag}
\newcommand{\LiZero}{\psi_{L,0}}

\newcommand{\bLDA}{\hat{b}}
\newcommand{\aNoo}{a^{N}_{11} }
\newcommand{\aNoz}{a^{N}_{10} }
\newcommand{\aNzz}{a^{N}_{00} }
\newcommand{\aLoo}{a^{L}_{11} }
\newcommand{\aLoz}{a^{L}_{10} }
\newcommand{\aLzz}{a^{L}_{00} }

\begin{document}

	\author{Alexander Mil}
	\affiliation{Universit\"{a}t Heidelberg, Kirchhoff-Institut f\"{u}r Physik,	Im Neuenheimer Feld 227, 69120 Heidelberg, Germany}
	\author{Torsten V.\ Zache}
	\affiliation{Universit\"{a}t Heidelberg, Institut f\"{u}r Theoretische Physik, Philosophenweg 16, 69120 Heidelberg, Germany}
	\author{Apoorva Hegde}
	\affiliation{Universit\"{a}t Heidelberg, Kirchhoff-Institut f\"{u}r Physik,	Im Neuenheimer Feld 227, 69120 Heidelberg, Germany}
	\author{Andy Xia}
	\affiliation{Universit\"{a}t Heidelberg, Kirchhoff-Institut f\"{u}r Physik,	Im Neuenheimer Feld 227, 69120 Heidelberg, Germany}
	\author{Rohit P.\ Bhatt}
	\affiliation{Universit\"{a}t Heidelberg, Kirchhoff-Institut f\"{u}r Physik,	Im Neuenheimer Feld 227, 69120 Heidelberg, Germany}
	\author{Markus K.\ Oberthaler}
	\affiliation{Universit\"{a}t Heidelberg, Kirchhoff-Institut f\"{u}r Physik,	Im Neuenheimer Feld 227, 69120 Heidelberg, Germany}
	\author{Philipp Hauke}
	\affiliation{Universit\"{a}t Heidelberg, Kirchhoff-Institut f\"{u}r Physik,	Im Neuenheimer Feld 227, 69120 Heidelberg, Germany}	
	\affiliation{Universit\"{a}t Heidelberg, Institut f\"{u}r Theoretische Physik, Philosophenweg 16, 69120
		Heidelberg, Germany}
	\author{J\"urgen Berges}
	\affiliation{Universit\"{a}t Heidelberg, Institut f\"{u}r Theoretische Physik, Philosophenweg 16, 69120 Heidelberg, Germany}
	\author{Fred Jendrzejewski}
	\affiliation{Universit\"{a}t Heidelberg, Kirchhoff-Institut f\"{u}r Physik,	Im Neuenheimer Feld 227, 69120 Heidelberg, Germany}
	
	\title{Realizing a scalable building block of a U(1) gauge theory with cold atomic mixtures}

	\date{\today}

	\begin{abstract}
		In the fundamental laws of physics, gauge fields mediate the interaction between charged particles. An example is quantum electrodynamics---the theory of electrons interacting with the electromagnetic field---based on U(1) gauge symmetry. Solving such gauge theories is in general a hard problem for classical computational techniques. While quantum computers suggest a way forward, it is difficult to build large-scale digital quantum devices required for complex simulations. Here, we propose a fully scalable analog quantum simulator of a U(1) gauge theory in one spatial dimension. To engineer the local gauge symmetry, we employ inter-species spin-changing collisions in an atomic mixture. We demonstrate the experimental realization of the elementary building block as a key step towards a platform for large-scale quantum simulations of continuous gauge theories.
	\end{abstract}
	
	\maketitle

	Continuous gauge symmetries are a cornerstone of our fundamental description of quantum physics as encoded in the Standard Model of Particle Physics.
	The presence of a gauge symmetry implies a concerted dynamics of matter and gauge fields that is subject to local symmetry constraints at each point in space and time~\cite{weinberg1995quantum}.
	To uncover the complex dynamical properties of such highly constrained quantum many-body systems requires enormous computational resources.
	This difficulty is stimulating great efforts to quantum simulate these systems, i.e., to solve their dynamics using highly controlled experimental setups with synthetic quantum systems~\cite{Wiese2013, Zohar2016,dalmonte2016lattice}. While first experimental breakthroughs have been achieved~\cite{Martinez2016,klco2018,Gorg2018,Schweizer2019}, the realization of scalable quantum simulators for gauge theories remains highly challenging.
	
	Our aim is the implementation of a continuous U(1) gauge theory, such as realized in quantum electrodynamics, in a scalable and highly tunable platform. 
	We base our setup on inter-species spin-changing collisions in an atomic mixture and propose an extended implementation scheme where the spin-changing collisions are isolated in single wells of an optical lattice. Thus, each well constitutes an elementary building block that includes the gauge-invariant interaction between matter and gauge fields. 
	This configuration decisively improves the involved time scales as compared to previous proposals, where spin-changing collisions had to be accompanied by hopping across different sites of the optical lattice~\cite{zohar2013quantum,Stannigel2013,Kasper2016,zache2018quantum}. 
	
	Based on this modular approach, we experimentally demonstrate the engineering of the elementary building block, demonstrate its high tunability, and verify its faithful representation of the desired model. 
	Since repetitions of this elementary unit can be connected using laser-assisted tunneling, our results pave the way for future setups to flexibly realize extended gauge systems. 
	
	\begin{figure}[!ht]
		\includegraphics[width=\columnwidth]{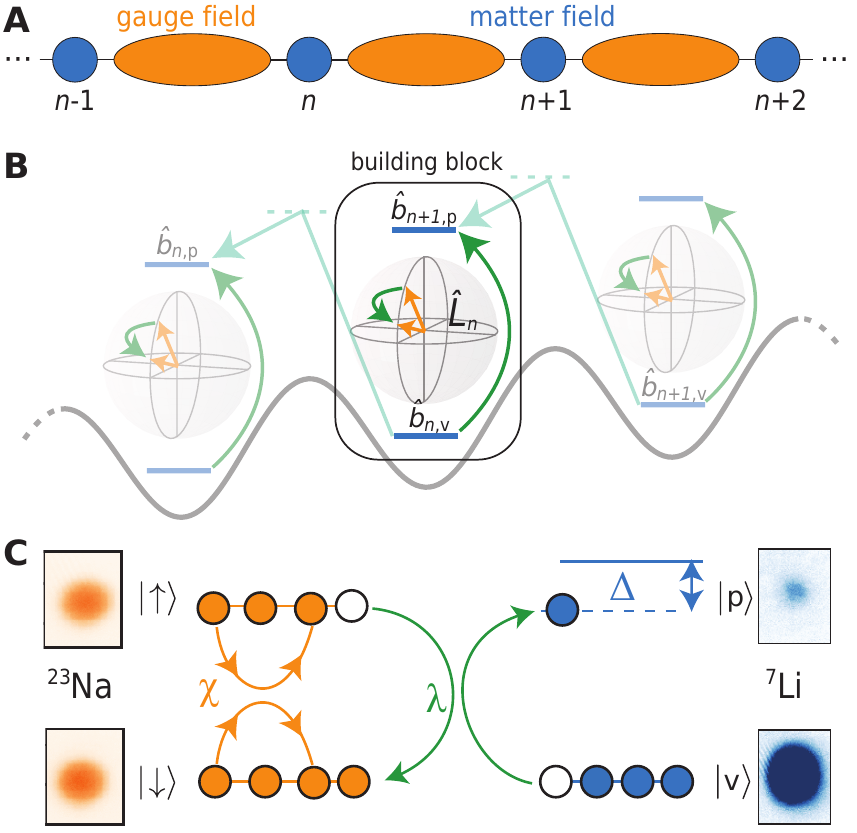}
		\caption{\label{fig:sketch}\textbf{Engineering a gauge theory.}
			(\textbf{A}) Structure of a lattice gauge theory. Matter fields reside on sites, gauge fields on the links in-between.
			(\textbf{B}) Proposed implementation of the extended system. Individual building blocks consist of long spins (representing gauge fields) and matter states, whose interaction constitutes a local U(1) symmetry. An array of building blocks in an optical lattice is connected via laser-assisted tunneling.
			(\textbf{C}) Experimental realization of the elementary building block. Gauge (sodium) and matter (lithium) states reside in a single well. The gauge-invariant interaction is realized by heteronuclear spin-changing collisions.
		}
	\end{figure}

	Scalability is an important ingredient for realistic applications to gauge field theory problems. Digital quantum simulations of gauge theories based on a Trotterized time evolution on a universal quantum computer, such as first realized using trapped ions~\cite{Martinez2016} or with superconducting qubits~\cite{klco2018}, are challenging to scale up. This difficulty makes analog quantum simulators, as treated here, highly attractive, since they can be scaled up while still maintaining excellent quantum coherence~\cite{bloch2008many,cirac2012goals,bernien2017probing,surace2019lattice,Kokail2019,berges2019scalingup}.
	
	In the past years, ultracold atoms have become a well-established platform for mimicking condensed-matter systems with external classical electric and magnetic fields~\cite{Goldman2014}.
	Techniques have been demonstrated for imbuing these fields with their own quantum dynamics by Floquet engineering in single-species atomic gases  \cite{Clark2018, Gorg2018}, and a first experiment has aimed at the realization of a discrete ($Z_2$) gauge symmetry in a minimal model~\cite{Schweizer2019}.
	
	At the same time, advances in cold atomic mixtures offer an exciting perspective, since mixtures naturally realize dynamical background fields for moving particles, as has been demonstrated in experiments on polaron phenomena \cite{Chevy2010, Grusdt2016}
	and the phononic Lamb shift \cite{Rentrop2016}. These systems possess \emph{global} U(1) symmetries related to the conservation of total magnetization and atom number~\cite{Stamper-Kurn2013}. However, a gauge theory is based on a \emph{local} symmetry. Our implementation scheme restricts the dynamics -- such that all corresponding local constraints in space and time are respected -- through a combination of spin-changing collisions in atomic mixtures and optical-lattice engineering. 
	
	We specify our proposal for a one-dimensional gauge theory on a spatial lattice, as visualized in Fig.\ \ref{fig:sketch}A. Charged matter fields reside on the lattice sites $n$, with gauge fields on the links in-between the sites \cite{kogut1979introduction}. We consider two-component matter fields labeled `p' and `v', which are described by the operators $(\hat{b}_{n,\text{p}},\hat{b}_{n,\text{v}})$. 
	To realize the gauge fields with the atomic system, we employ the quantum link formulation \cite{Horn1981, Chandrasekharan1997, Banerjee2012}, where the gauge fields are replaced by quantum mechanical spins $\Lop_n$, labeling link operators by the index of the site to the left. In this formulation, the spin $z$-component $\hat{L}_{n,z}$ can be identified with a discrete `electric' field. We recover the continuous gauge fields of the original quantum field theory in a controlled way by working in the limit of long spins~\cite{Kasper2016}.
	
	Physically, this system of charged matter and gauge fields can be realized in a mixture of two atomic Bose--Einstein condensates (BECs) with two internal components each (in our experiment, we use \textsuperscript{7}Li and \textsuperscript{23}Na).
	An extended system can be obtained by use of an optical lattice. In our scheme, we abandon the one-to-one correspondence between the sites of the simulated lattice gauge theory and the sites of the optical-lattice simulator. This correspondence characterized previous proposals and necessitated physically placing the gauge fields in-between matter sites~\cite{zohar2013quantum,Stannigel2013,Kasper2016,zache2018quantum}.
	Instead, as illustrated in Fig.~\ref{fig:sketch}B, here one site of the physical lattice hosts two matter components, each taken from one adjacent site ($\hat{b}_{n+1,\text{p}}$ and $\hat{b}_{n,\text{v}}$), as well as the link ($\Lop_n$).
	
	The enhanced physical overlap in this configuration decisively improves time scales of the spin-changing collisions, which up to now was a major limiting factor for experimental implementations.
	Moreover, a single well of the optical lattice already contains the essential processes between matter and gauge fields, and thus represents an elementary building block of the lattice gauge theory.
	These building blocks can be coupled by laser-assisted tunneling of the matter fields.
	
	The Hamiltonian $\hat{H} =\sum_n [\hat{H}_n + \hbar\Omega (\hat{b}_{n,\mathrm{p}}^\dagger\hat{b}_{n,\mathrm{v}}+\mathrm{h.c.})]$ of the extended system can thus be decomposed into the elementary building-block Hamiltonian $\hat{H}_n$ and the laser-assisted tunneling (with Raman frequency $\sim \Omega$).
	Here, $\hat{H}_n$ reads (writing $\bMFzero\equiv\hat{b}_{n+1,\text{p}}$, $\bMFone\equiv\hat{b}_{n,\text{v}}$, and $\Lop\equiv\Lop_n$)
	\begin{align}
	\hat{H}_n / \hbar &= \chi\Lz^2 + \frac{ \Delta}{2} \left(\bMFzero^\dagger \bMFzero  -  \bMFone^\dag \bMFone\right) \nonumber\\
	&+  \lambda \left(\hat{b}_\text{p}^\dag \Lmin \hat{b}_\text{v} + \hat{b}_\text{v}^\dag \Lpl \hat{b}_\text{p} \right).\label{eq:Hamiltonian}
	\end{align}
	The first term on the right-hand side of Eq.~(\ref{eq:Hamiltonian}), which is proportional to the parameter $\chi$, describes the energy of the gauge field, while the second term $\sim\Delta$ sets the energy difference between the two matter components. The last term $\sim \lambda$ describes the $U(1)$ invariant coupling between matter and gauge fields, which is essential to retain the local $U(1)$ gauge symmetry of the Hamiltonian $\hat{H}$ (see SM for more details).
	
	We implement the elementary building block Hamiltonian $\hat{H}_n$ with a mixture of $\num{300e3}$ sodium and $\num{50e3}$ lithium atoms as sketched in Fig.~\ref{fig:sketch}C (see SM for details). Both species are kept in an optical dipole trap such that the external trapping potential is spin-insensitive for both species. An external magnetic bias field of $B \approx \SI{2}{G}$ suppresses any spin change energetically, such that only the two Zeeman levels, $m_F = 0$ and $1$, of the $F=1$ hyperfine ground state manifolds are populated during the experiment. The \textsuperscript{23}Na states are labelled as $\ket{\uparrow}=\ket{m_F=0}$ and $\ket{\downarrow}=\ket{m_F=1}$, on which the spin operator $\Lop$ associated to the gauge field acts. The first term of \eqref{eq:Hamiltonian} is then identified with the one-axis twisting Hamiltonian~\cite{Gross2010,riedel2010atom}.
	We label the \textsuperscript{7}Li states as `particle' $\ket{\text{p}}=\ket{m_F=0}$ and `vacuum' $\ket{\text{v}}=\ket{m_F=1}$, in accordance with the matter field operators $\bMFzero$ and $\bMFone$. With this identification, the second term arises from energy shifts due to the external magnetic field and density interactions.
	Finally, the term $\sim\lambda$ is physically implemented by heteronuclear spin-changing interactions~\cite{XuPRL}.

	\begin{figure}
		\includegraphics[width=\columnwidth]{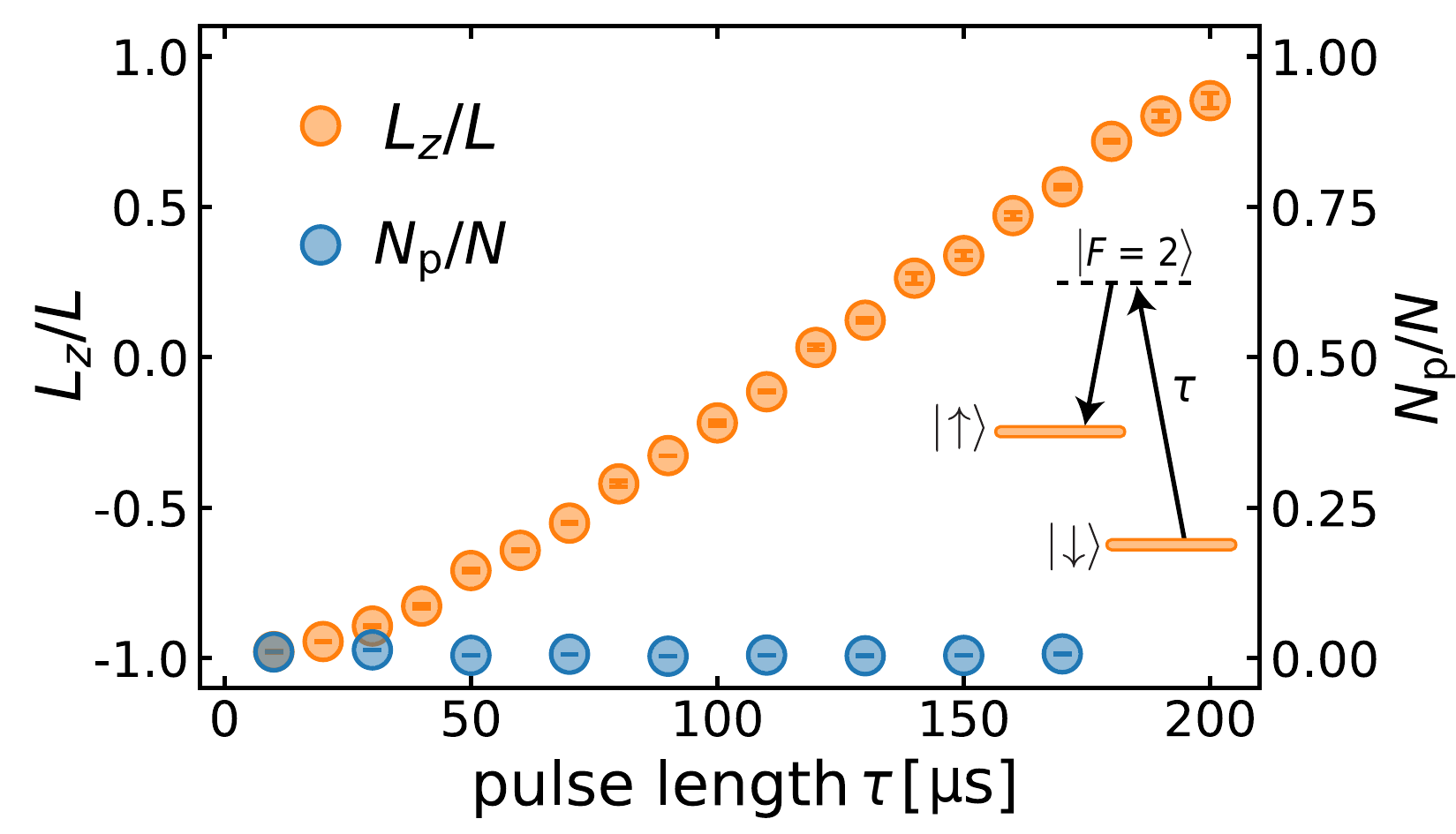}
		\caption{\label{fig:initialconditions}\textbf{Tunability of the initial conditions.}
			The normalized spin z-component $L_z/L$ of \textsuperscript{23}Na atoms as a function of the preparation pulse length, which shows that the gauge field can be tuned experimentally over the entire possible range. Simultaneously, the particle number $N_\text{p}/N$ of \textsuperscript{7}Li is kept in the vacuum state. The inset shows a sketch of the experimental protocol used for tuning the initial conditions.
		}
	\end{figure}
	
	The resulting setup is highly tunable,
	as we now demonstrate experimentally on the building block. We achieve tunability of the gauge field through a two-pulse Rabi coupling of the Na atoms between $\ket{\downarrow}$ and  $\ket{\uparrow}$ using an intermediate $\ket{F=2}$ state, which yields a desired value of $L_z/L = (\nNup-\nNdown)/(\nNup+\nNdown)$ (Fig.~\ref{fig:initialconditions}). At the same time, we keep the \textsuperscript{7}Li atoms in $\ket{\text{v}}$, corresponding to the initial vacuum of the matter sector at $\Delta\to\infty$, with $\lesssim 1\%$ detected in $\ket{\text{p}}$ (see Fig.~\ref{fig:initialconditions}).
	
	\begin{figure}
		\includegraphics[width=0.95\columnwidth]{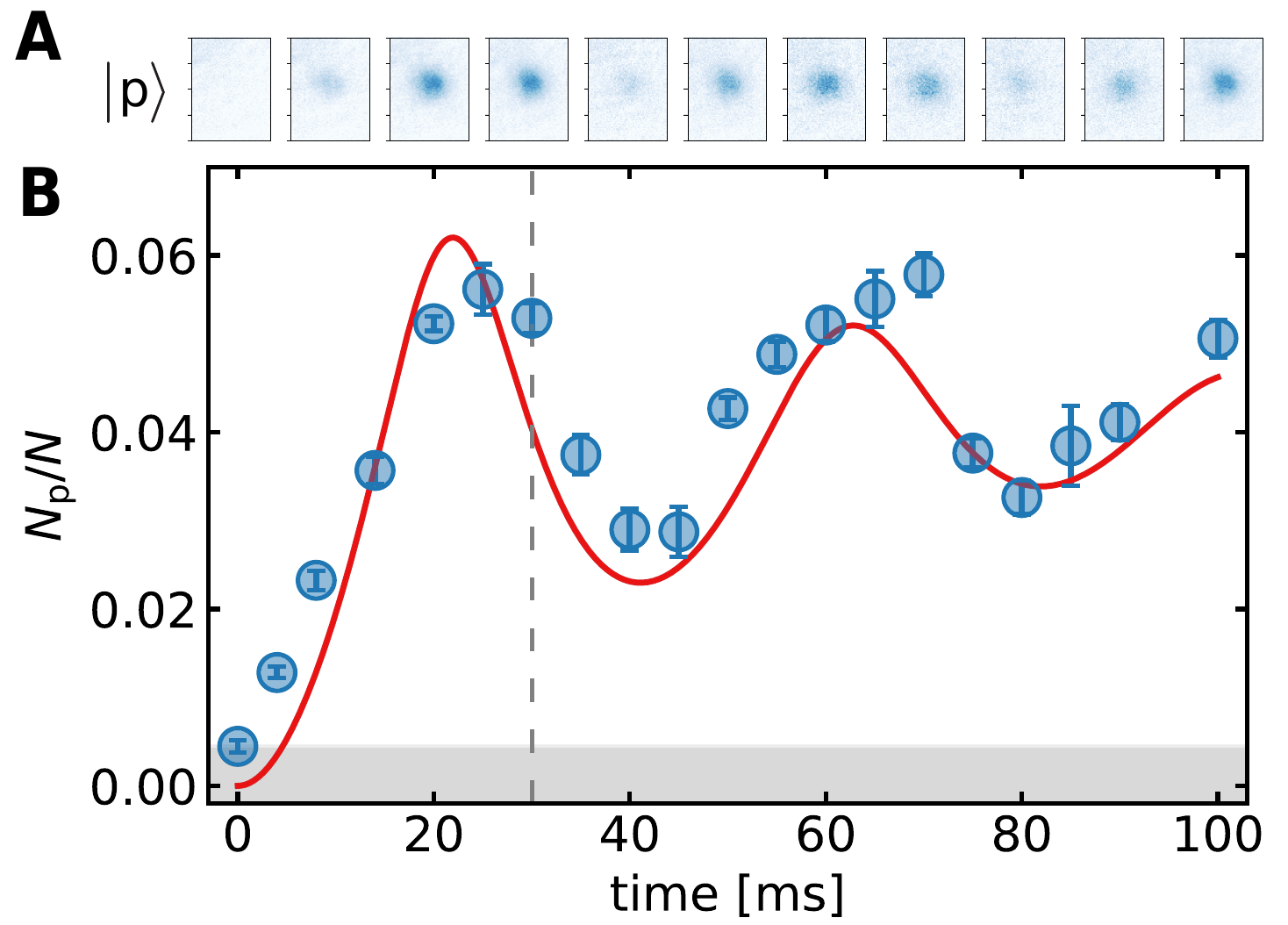}
		\caption{\label{fig:gaugefielddynamics}\textbf{Dynamics of particle production.} (\textbf{A}) The number density distribution in state $\ket{\text{p}}$ as a function of time for $\fGauge=-0.188$. (\textbf{B}) The corresponding particle number $N_\text{p}/N$. The blue circles give the experimental values with bars indicating the statistical error on the mean. The red curve is the theoretical mean-field prediction of Hamiltonian (\ref{eq:Hamiltonian}) with parameters determined from a fit of the data in Fig.~\ref{fig:gaugefieldB}A and phenomenological damping. The shaded area indicates the experimental noise floor. The dashed line marks the time of 30ms as it is used for the experimental sequence generating the data in Fig.~\ref{fig:gaugefieldB}.}
	\end{figure}
	
	If the gauge invariant coupling is turned off by removing the Na atoms from the trap, we observe no dynamics in the matter sector beyond the detection noise. On the other hand, once
	the gauge field is present, the matter sector clearly undergoes a transfer from $\ket{\text{v}}$ to $\ket{\text{p}}$ for proper initial conditions, as illustrated in Fig.~\ref{fig:gaugefielddynamics}A for an initialization to $\fGauge=-0.188$ at a magnetic field of $B_\mathrm{A} = \SI{2.118(2)}{G}$~\footnote{All uncertainties given in this manuscript correspond to a 68\% confidence interval.}.
	This observation demonstrates the controlled operation of heteronuclear spin-changing collisions implementing the gauge-invariant dynamics in the experiment.
	
	To quantify our observations, we extract the ratio $\fParticle$, with $N=N_\text{p} + N_\text{v}$, as a function of time as presented in Fig.~\ref{fig:gaugefielddynamics}B.
	We observe non-zero $\fParticle$, describing `particle production', on a timescale of few tens of $\SI{}{\milli \second}$, with up to $6\%$ of the total $N$ being transferred to  $\ket{\text{p}}$. This value is consistent with our expectations from conservation of the initial energy $E_0/\hbar = \chi L_{z}^2 - \Delta N_\text{v}/2$, from which we estimate a maximum amplitude on the order of a few percent.
	Due to the much larger \textsuperscript{23}Na condensate, the expected corresponding change in $\fGauge$ is $\sim 2\%$, which is currently not detectable with our imaging routine (see SM for details). Coherent oscillations in $\fParticle$ are seen to persist for about $\SI{100}{\milli \second}$.  
	
\begin{figure}
		\includegraphics[width=\columnwidth]{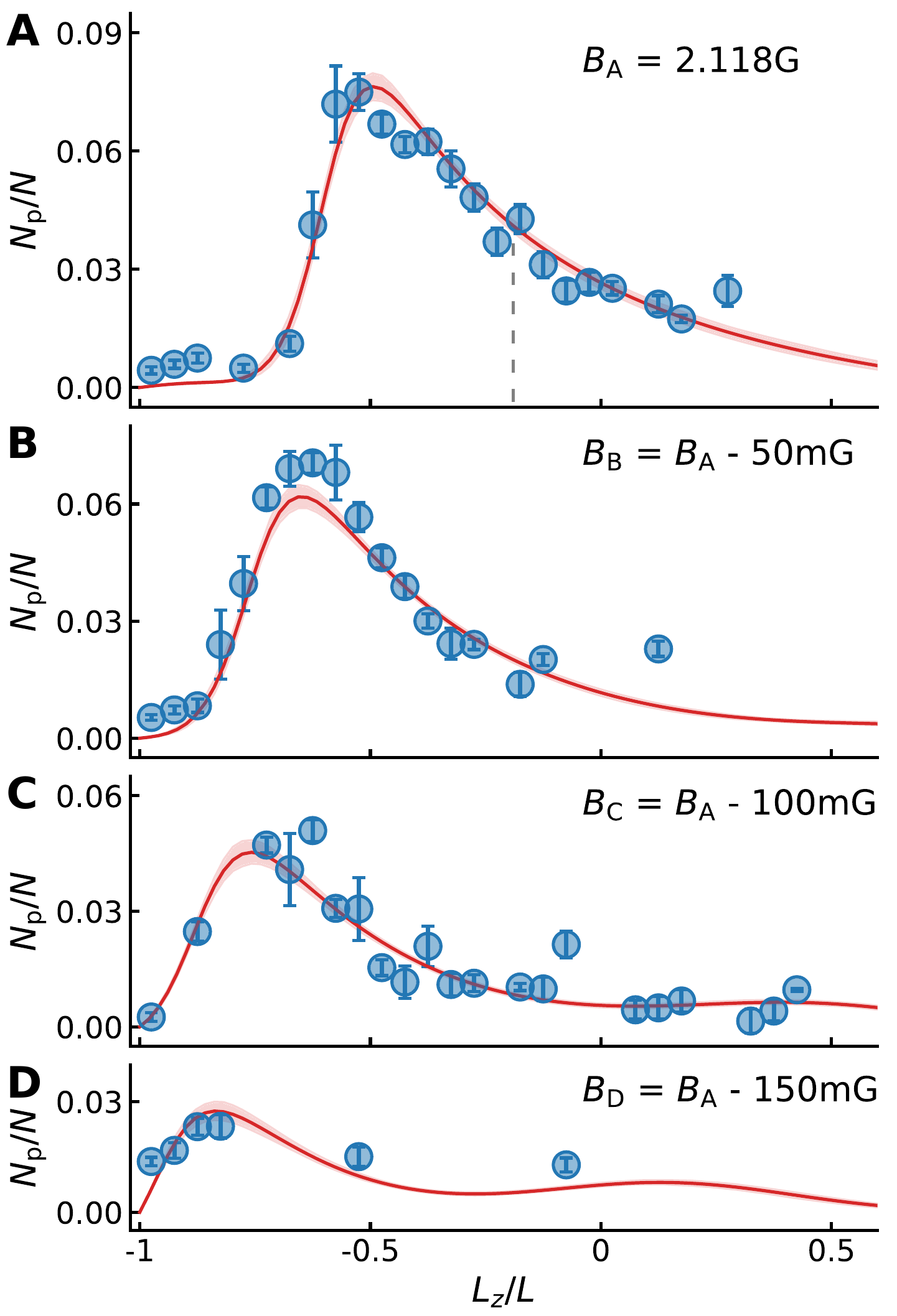}
		\caption{\label{fig:gaugefieldB}\textbf{Resonant particle production.} (\textbf{A}-\textbf{D}) The number of produced particles as a function of initially prepared $L_z/L$ after 30ms for different bias magnetic fields. Blue circles are experimental values with bars indicating the error on the mean.
			The red curve in (A) arises from the theoretical model using the best estimate values of $\chi$, $\lambda$, $\Delta_0$ and $\Delta_L$. The remaining curves in (B-D) are computed using the same parameters including $\Delta_B$. The shaded area indicates confidence intervals of the fit from bootstrap resampling.
			The dashed line in (A) indicates the $L_z/L$ value corresponding to the time evolution shown in Fig.~\ref{fig:gaugefielddynamics}.
		}
	\end{figure}

	We display $N_P/N$ over the entire range of initial $L_z$ in Fig.~\ref{fig:gaugefieldB}, keeping a fixed time of $\SI{30}{\milli \second}$. The upper panel A corresponds to the same experimental setting as in Fig.~\ref {fig:gaugefielddynamics}. A clear resonance for particle production can be seen around $L_z/L \simeq -0.5$, approximately captured by the resonance condition $2\chi L_z\sim\Delta$ (see SM for details). The asymmetry of the resonance is a clear manifestation of the nonlinearity of the dynamics. As we reduce the magnetic field $B$, presented in Fig.~\ref{fig:gaugefieldB}B-D, we observe a shift of the resonant particle production together with a reduction in amplitude, which continues to be qualitatively captured by the resonance condition $2\chi L_z \sim \Delta$. The maximal amplitude of the particle production is necessarily reduced by the conservation of total magnetization as the resonant peak is pushed closer to $L_z / L=-1$.
	For fields that are smaller than $B_\textrm{min} \approx \SI{1.96}{G}$, the matter field and gauge field dynamics become too off-resonant and particle production can no longer be observed.
	
	We compare the experimental results to the mean-field predictions of Hamiltonian (\ref{eq:Hamiltonian}) for chosen $\chi$, $\lambda$ and $\Delta(L_z,B) = \Delta_0 + \Delta_L L_z/L + \Delta_B (B-B_A)/B_A$ (see SM for the origin of the dependence on the magnetic field $B$ and the initial spin $L_z$).
	A first-principle calculation of these model parameters, using only experimental input of our setup, yields
	$\chi^\text{th} / 2\pi \approx \SI{21.42}{\milli \hertz}$, $\lambda^\text{th}/ 2\pi \approx \SI{123.7}{\micro\Hz}$, $\Delta^\text{th}_0 / 2\pi \approx -\SI{31}{\Hz}$, $\Delta^\text{th}_L / 2\pi \approx  \SI{6.546}{\kilo\Hz}$ and $\Delta^\text{th}_B/ 2\pi \approx -\SI{1.669}{\kilo\Hz}$.
	These values are obtained by neglecting any residual spatial dynamics~\cite{nicklas2011rabi} of the atomic clouds within the trapping potential, which renormalizes the model parameters.  
	Moreover, the mean-field approximation cannot capture the decoherence observed in Fig.~\ref {fig:gaugefielddynamics} at later times. However, the features of the resonance data in Fig.~\ref{fig:gaugefieldB} are more robust against the decoherence as it probes the initial rise of particle production. 
	
	We include the decoherence into the model phenomenologically by implementing a damping term characterized by $\gamma/ 2\pi = \SI{3.54(94)}{\hertz}$, which is determined by an exponential envelope fit to the data of Fig.~\ref{fig:gaugefielddynamics}B. Fixing $\gamma$, the best agreement (solid red line) with the data in Fig.~\ref{fig:gaugefieldB}A is obtained for
	$\chi / 2\pi = \SI{8.802(8)}{\milli\Hz}$, $\lambda /2\pi =  \SI{16.4(6)}{\micro\Hz}$, $\Delta_0 / 2\pi = -\SI{4.8(16)}{\Hz}$ and $\Delta_L /2\pi =  \SI{2.681(1)}{\kilo\Hz}$. The prediction with these model parameters shows excellent agreement with the data in Fig.~\ref{fig:gaugefielddynamics}B (red line) for all times observed. Remarkably, our established model also describes the data in Fig.~\ref{fig:gaugefieldB}B-D by including $\Delta_B / 2\pi = -\SI{519.3(3)}{ \hertz}$ (see SM). Compared to the ab-initio estimates, all fitted values have the expected sign and lie in the same order of magnitude.

	Our results demonstrate the controlled operation of an elementary building block of a $U(1)$ gauge theory, and thus open the door for large-scale implementations of lattice gauge theories in atomic mixtures.
	The resulting extended gauge theory will enable the observation of relevant phenomena such as plasma oscillations or resonant particle production in strong-field QED~\cite{Kasper2014}.
	Along the path to the relativistic gauge theories realized in nature, we will replace bosonic \textsuperscript{7}Li with fermionic \textsuperscript{6}Li, which will allow for the recovery of Lorentz-invariance in the continuum limit.
	
	\begin{acknowledgments}
		This work is part of and supported by the DFG Collaborative Research Centre ``SFB 1225 (ISOQUANT)'', the ERC Advanced Grant ``EntangleGen'' (Project-ID 694561),
		the ERC Starting Grant ``StrEnQTh'' (Project-ID 804305), and
		the Excellence Initiative of the German federal government and the state governments – funding line Institutional Strategy (Zukunftskonzept): DFG project number ZUK 49/Ü. F. J. acknowledges the DFG support through the project FOR 2724, the Emmy-Noether grant (project-id 377616843) and from the Juniorprofessorenprogramm Baden-W\"urttemberg (MWK). 
	\end{acknowledgments}

%

\clearpage
\newgeometry{a4paper,
	left=0.75in,right=0.483in,top=0.75in,bottom=1.55in
}

\onecolumngrid
~ ~
\begin{center} 
	\begin{Large} \textbf{Supplementary Material} \end{Large}
\end{center}
\medskip
\smallskip
\twocolumngrid

\setcounter{section}{0}
\setcounter{subsection}{0}
\setcounter{figure}{0}
\setcounter{equation}{0}
\setcounter{NAT@ctr}{0}

\renewcommand{\figurename}[1]{FIG. }
\renewcommand{\theequation}{S\arabic{equation}}

\makeatletter
\renewcommand{\bibnumfmt}[1]{[S#1]}
\renewcommand{\citenumfont}[1]{S#1}
\renewcommand{\thefigure}{S\@arabic\c@figure} 
\makeatother
\renewcommand{\theequation}{S\arabic{equation}} 
\renewcommand\thetable{S\arabic{table}}
\renewcommand{\vec}[1]{{\boldsymbol{#1}}}

\section{Experimental implementation}\label{App:ExpertimentalDetails}

We prepare a mixture of bosonic \textsuperscript{23}Na and \textsuperscript{7}Li in a single crossed optical dipole trap from a far red detuned laser with \SI{1064}{\nano\meter} wavelength. The atoms experience trapping frequencies of ($\omega_x,\omega_y,\omega_z)_{Na} \approx 2\pi\times (243,180,410)$Hz for sodium and ($\omega_x,\omega_y,\omega_z)_{Li} = 2.08 \times (\omega_x,\omega_y,\omega_z)_{Na}$ for lithium. To maximize the spatial overlap of the two individual clouds, we align the direction of strongest confinement in gravity direction whereby reducing the differential gravitational sag to approximately 1.1$\mu m$. The atomic clouds are evaporatively cooled to Bose Einstein condensation and contain about 300 $\times 10^3$ \textsuperscript{23}Na atoms and 50 $\times 10^3$ \textsuperscript{7}Li atoms respectively.

For the spin exchange dynamics we take into account the following sublevels of the Hyperfine groundstate: $\ket{\downarrow} = \ket{F = 1, m_F = 1}, \ket{\uparrow} = \ket{F = 1, m_F = 0}$ for sodium, as well as $\ket{v} = \ket{F = 1, m_F = 1}, \ket{p} = \ket{F = 1, m_F = 0}$ for lithium.
We apply an offset magnetic field of $B_A = 2.118(2)\mathrm{G}$, which lifts the degeneracy of the magnetically sensitive states. Moreover around B$_0$ the level spacing of both species approach each other at $(E_{\ket{\uparrow}} - E_{\ket{\downarrow}})/\hbar \approx (E_{\ket{\mathrm{p}}} - E_{\ket{\mathrm{v}}})/\hbar \sim \SI{1.45}{\mega \Hz}$, energetically allowing heteronuclear spin transfer. Detailed information about the scattering lengths in our system is given in section B.

At the beginning of the experimental sequence the atoms are prepared in $\ket{\downarrow}$ and  $\ket{v}$ respectively, supressing any spin exchange due to conservation of magnetization. We initiate the spin dynamics by quenching sodium into a desired superposition of $\ket{\uparrow}$ and $\ket{\downarrow}$. To ensure that lithium stays spin polarized during the quench, instead of direct radiofrequency transfer we couple the sodium states with a two pulse microwave transition sequence using the sodium$\ket{F =2, m_F = 0}$ state as an intermediate state. The first pulse is of variable length, driving population from $\ket{\downarrow}$ to the intermediate state with a Rabi frequency of $2\pi\times \SI{2.5}{\kilo \hertz}$. The second pulse is fixed to $\SI{100}{\mu\second}$ (corresponding to a $\pi$-pulse) and subsequently drives the population from the intermediate state to $\ket{\uparrow}$. The total length of the pulse sequence is at most $\SI{300}{\mu\second}$, which is at least one order of magnitude faster than the spin dynamics we observe in our experiment.
\renewcommand{\citenumfont}[1]{#1}
The initial superposition quench causes undesired external dynamics of the sodium cloud due to the two states $\ket{\uparrow}$ and $\ket{\downarrow}$ being immiscible~\cite{nicklas2011rabi}. 
\renewcommand{\citenumfont}[1]{S#1}
This leads to relative motion of the two sodium spin components within the trap after the superposition quench. Our observation time is limited by losses in the $\ket{\uparrow}$ state, leading to a lifetime  of the $\ket{\uparrow}$ population of about \SI{560}{\milli \second} for the data shown in Fig.~3.

After an evolution time of up to \SI{100}{\milli \s} we switch off the trapping potential and perform a Stern Gerlach sequence, where a magnetic field gradient is applied across the atom's position for \SI{1.5}{\milli\second} that separates the magnetic substates spatially. After a free expansion time of \SI{3}{\milli \s} (sodium) and \SI{2}{\milli \s} (lithium) the two spin states of each species are separated by approximately \SI{130}{\mu \meter}. Then we use absorption imaging to detect the spatial distribution of both species on two individual CCD cameras. The information about the population of substates is then extracted from the images by integrating the density distribution of the individual spin components.

The observable in the experiment is the relative lithium atom number population $N_\mathrm{p}/N$. For lithium being initially spin polarized in $\ket{\mathrm{v}}$, atoms transferred to $\ket{\mathrm{p}}$ are the signature of spin changing collisions. The main source of detection noise for this observable are fringe patterns on the images which are resulting from diffraction and interference in the imaging path. To account for those we postprocess the lithium absorption images by applying a fringe removal algorithm~\cite{Ockeloen2010fringe_removal}, which reduces the experimental noise floor below 1$\%$ (see ~Fig. 3B).

Due to the 6 times higher sodium atom number the relative change between the population of $\ket{\uparrow}$ and $\ket{\downarrow}$ resulting from spin changing collisions is expected to be at most 1$\%$. This change can currently not be detected with our imaging routine. Main limitation are the external dynamics of the sodium spin components in the trap after the initial quench. This results in strong variation of the density distribution of the individual sodium components which our imaging calibration is systematically sensitive to \cite{Reinaudi2007imaging}. 

The observable $N_\mathrm{p}/N$ is obtained as an average from a statistical ensemble of multiple experimental realizations. Each data point shown in Figs.~3 and 4 corresponds to at least three and ten measurements, respectively. The data displayed in Fig.~4 is binned in $5\%$ intervals of $L_z/L$.

\section{Microscopic Hamiltonian}
In our experiments, we work with two internal spin states, $m_F = 0,1$, each from the spin $F_s = 1$ manifold
of the two Bose gases, $s = N, L$ (${}^{23}\text{Na}$ and ${}^{7}\text{Li}$ in our case). The total Hamiltonian, $H = H_0 + H_1$, of the combined system splits into free, $H_0 = H_N + H_L$, and interaction, $H_1 = H_{NN} + H_{LL} + H_{NL}$, parts~\cite{luo2007bose}\renewcommand{\citenumfont}[1]{#1}~\cite{Stamper-Kurn2013}\renewcommand{\citenumfont}[1]{S#1}. The free parts have the form
\begin{subequations}
	\begin{align}\label{eq:free_part}
	\hat{H}_s &= \int d^3\mathbf{x} \sum_\alpha \hat{\psi}_{s,\alpha}^\dagger(\mathbf{x}) \hat{\mathcal{H}}_{s,\alpha}\left(\mathbf{x}\right)\hat{\psi}_{s,\alpha}\left(\mathbf{x}\right)\;, \\
	\hat{\mathcal{H}}_{s,\alpha}\left(\mathbf{x}\right) &= \frac{-\nabla_\mathbf{x}^2}{2m_s} + V_{s}(\mathbf{x}) + E_{s,\alpha}(B) \; .
	\end{align}
\end{subequations}
Here, $m_s$ denotes the atomic masses, $V_s$ is the trapping potential and $E_{s,\alpha}(B)$ is the Zeeman shift in the presence of an external magnetic field $B$, given by the Breit-Rabi formula. The field operators $\hat{\psi}_{s,\alpha}(\mathbf{x})$ fulfill bosonic commutation relations $\left[\hat{\psi}_{s,\alpha}(\mathbf{x}), \hat{\psi}_{s',\beta}^\dagger(\mathbf{y}) \right] = \delta_{s s'}\delta_{\alpha \beta} \delta \left(\mathbf{x}-\mathbf{y}\right)$,
where $\alpha,\beta \in \left\lbrace 0,1 \right\rbrace$ denote the $m_F$ states.

The intra-species and inter-species interactions which are spin-conserving are described by
\begin{subequations}
	\begin{align}
	\hat{H}_{ss} &= \frac{1}{2}\int d^3\mathbf{x} \sum_{\alpha, \beta} g_{\alpha \beta}^s \, \hat{\psi}_{s,\alpha}^\dagger (\mathbf{x}) \hat{\psi}_{s,\beta}^\dagger (\mathbf{x}) \hat{\psi}_{s,\beta} (\mathbf{x}) \hat{\psi}_{s,\alpha}(\mathbf{x}) \;, \\\hat{H}_{NL} &= \int d^3\mathbf{x}\sum_{\alpha, \beta} g_{\alpha \beta}^{Mix} \, \hat{\psi}_{N,\alpha}^\dagger (\mathbf{x}) \hat{\psi}_{L,\beta}^\dagger (\mathbf{x}) \hat{\psi}_{L,\beta} (\mathbf{x}) \hat{\psi}_{N,\alpha}  (\mathbf{x}) \;,
	\end{align}
\end{subequations}
\renewcommand{\citenumfont}[1]{#1}
where the interaction constants $g^s_{\alpha \beta} = \frac{4\pi\hbar^2}{m_s}a^s_{\alpha \beta}$ and $g^{Mix}_{\alpha \beta} = \frac{2\pi\hbar^2}{\mu}a^{Mix}_{\alpha \beta}$ are determined by the following scattering lengths: $\aNoo=\aNoz=a^N_{01}= 55\aBohr$, $\aNzz= 53\aBohr$, $\aLoo=\aLoz=a^L_{01} = 6.8\aBohr$, $\aLzz = 12.5\aBohr$~\cite{Stamper-Kurn2013} 
\renewcommand{\citenumfont}[1]{S#1}
and $a^{Mix}_{00} = a^{Mix}_{10} = a^{Mix}_{01}=19.65\aBohr$, $a^{Mix}_{11} = 20\aBohr$~\cite{Tiemann}. The hetero-nuclear spin-changing collisions are described by
\begin{align}\label{eq:inter_int_SCC}
\hat{H}_{SCC} &= \gSCC\int d^3\mathbf{x}  \, \dNaZero (\mathbf{x}) \dLiOne (\mathbf{x}) \NaOne (\mathbf{x}) \LiZero (\mathbf{x}) \nonumber\\
&+ \text{h.c.}
\end{align}
with the interaction strength $g^{SCC} = \frac{2\pi\hbar^2}{\mu}a_{SCC}$ and scattering length $a_{SCC} = 0.35\aBohr$~\cite{Tiemann}.

\section{Proposed extended gauge theory}
We propose to implement an extended $U(1)$ gauge theory in one spatial dimension as follows. Starting from the microscopic Hamiltonian a deep optical lattice localizes the atomic clouds on individual lattice wells $n$. Expanding the field operators into localized Wannier functions, $\hat{\psi}_{s,\alpha}=\sum_n \hat{b}_{s,\alpha,n} \Phi_{s,\alpha,n}(\mathbf{x})$, we first perform the spatial integration to obtain a tight-binding Hamiltonian. For a sufficiently deep potential, all tunneling and interaction terms between neighbouring wells are neglibible and resulting in an array of elementary building blocks $n$, each containing both species. The Hamiltonian has the form $\hat{H} = \sum_n \hat{H}_n$ with $\hat{H}_n = \sum_{s,n}\left(\hat{H}_{s,n}+ \hat{H}_{ss,n} \right) + \hat{H}_{NL,n} + \hat{H}_{SCC,n}$ and
\begin{widetext}
	\begin{subequations}
		\begin{align}
		\hat{H}_{s,n} &= \sum_{\alpha} E_{s,\alpha}(B) \bLDA_{s,\alpha, n}^{\dagger}\bLDA_{s,\alpha, n}\;,\\
		\hat{H}_{ss,n}&= X_{11}^s \bLDA_{s,1, n}^{\dagger}\bLDA_{s,1, n}^{\dagger}\bLDA_{s,1, n}\bLDA_{s,1, n}+ X_{00}^s \bLDA_{s,0, n}^{\dagger}\bLDA_{s,0, n}^{\dagger}\bLDA_{s,0, n}\bLDA_{s,0, n}+ 2 X_{10}^s  \bLDA_{s,1, n}^{\dagger}\bLDA_{s,0, n}^{\dagger}\bLDA_{s,1, n}\bLDA_{s,0, n}\;,\\
		\hat{H}_{NL,n} &= X^{Mix}_{11} \bLDA_{N,1, n}^{\dagger}\bLDA_{N,1, n}\bLDA_{L,1, n}^{\dagger}\bLDA_{L,1, n} + X^{Mix}_{00} \bLDA_{N,0, n}^{\dagger}\bLDA_{N,0, n}\bLDA_{L,0, n}^{\dagger}\bLDA_{L,0, n} \\
		&+ X^{Mix}_{10} \bLDA_{N,1, n}^{\dagger}\bLDA_{N,1, n}\bLDA_{L,0, n}^{\dagger}\bLDA_{L,0, n} +X^{Mix}_{10} \bLDA_{N,0, n}^{\dagger}\bLDA_{N,0, n} \bLDA_{L,1, n}^{\dagger}\bLDA_{L,1, n} \;,\\
		\hat{H}_{SCC,n}&= X^{SCC} \left[ \bLDA_{N,0, n}^{\dagger} \bLDA_{L,1, n}^{\dagger} \bLDA_{N,1, n} \bLDA_{L,0, n} + \bLDA_{N,1, n}^{\dagger} \bLDA_{L,0, n}^{\dagger} \bLDA_{N,0, n} \bLDA_{L,1, n}\right] \;.
		\end{align}
	\end{subequations}
\end{widetext}
The precise values of the energy levels $E$ and interaction constants $X$ depend on the details of the optical lattice and the corresponding Wannier functions $\Phi_{s,\alpha,n}(\mathbf{x})$.

We then identify the gauge fields on the building block via the Schwinger representation of angular momentum operators,
\begin{subequations}
	\begin{align}
	\hat{L}_{+,n} &= \hat{b}_{N,0,n}^\dagger \hat{b}_{N,1,n} \;, \\ \hat{L}_{-,n} &= \hat{b}_{N,1,n}^\dagger \hat{b}_{N,0,n} \;, \\
	\hat{L}_{z,n} &= \frac{1}{2}\left(\hat{b}_{N,0,n}^\dagger \hat{b}_{N,0,n} - \hat{b}_{N,1,n}^\dagger \hat{b}_{N,1,n}\right) \;.
	\end{align}
\end{subequations}
The matter fields on a building block, however, are partially associated with the two neighbouring internal degrees of freedom according to
\begin{align}
\hat{b}_n = \begin{pmatrix}
\hat{b}_{n,\text{p}}  \\ \hat{b}_{n,\text{v}}
\end{pmatrix} \;, && \hat{b}_{n,\text{p}} = \hat{b}_{L,n-1,0}  \;, && \hat{b}_{n,\text{v}} = \hat{b}_{L,n,1}  \;.
\end{align}
The matter fields carry a $U(1)$ charge $\hat{Q}_n$, which enters the Gauß' law operators $\hat{G}_n$ as
\begin{subequations}
	\begin{align}
	\hat{Q}_n &= \hat{b}_n \hat{b}_n = \hat{b}_{n,\text{p}}^\dagger \hat{b}_{n,\text{p}} + \hat{b}_{n,\text{v}}^\dagger \hat{b}_{n,\text{v}} \;, \\
	\hat{G}_n &= \hat{L}_{z,n} - \hat{L}_{z,n-1} - \hat{Q}_n \;.
	\end{align}
\end{subequations}
This construction ensures Gauß' law, $\left[\hat{G}_n,\hat{H}\right] =  \left[\hat{G}_n,\hat{H}_{SCC,n}+ \hat{H}_{SCC,n-1}\right]= 0$, because
$\left[\hat{L}_{z,n}, \hat{H}_{SCC,n}\right] - \left[\hat{L}_{z,n-1}, \hat{H}_{SCC,n-1}\right] = \left[\hat{Q}_n,\hat{H}_{SCC,n}+ \hat{H}_{SCC,n-1}\right]$.
Physically, a global $U(1)$ symmetry due to the conservation of the total magnetization, which arises from angular momentum conservation, is localized on each building block.
This localization, which is achieved by associating a building block $n$ with a link $n$ and components of the matter fields on the neighbouring lattice sites $n$ and $n+1$, is the key to the scalability of our proposal.

Finally, we connect the building blocks to an extended one-dimensional system by the term
\begin{align}
H_\Omega  &= \hbar \Omega	\sum_n  \left(\hat{b}_{L,n,1}^\dagger \hat{b}_{L,n-1,0} + \text{h.c.}\right)  \nonumber \\ &= \hbar \Omega\sum_n  \left(\hat{b}_{n,\text{v}}^\dagger \hat{b}_{n,\text{p}} + \text{h.c.}\right)  \;,
\end{align}
which can be realized, e.g., with laser-assisted tunneling. Crucially, this connection respects Gauß' law, $\left[H_\Omega, \hat{G}_n\right]=\left[H_\Omega, \hat{Q}_n\right]=0$.  
\renewcommand{\citenumfont}[1]{#1}
The resulting lattice model is similar to our previous proposal~\cite{zache2018quantum} with Dirac fermions replaced by two-component bosons,
\renewcommand{\citenumfont}[1]{S#1}
\begin{align}\label{eq:extended_gauge_theory}
\hat{H}/\hbar &= \sum_n \left\lbrace\chi \; \hat{L}_{z,n}^2 + \Omega \; \hat{b}_n^\dagger \sigma_x\hat{b}_n  \right. \\
&\qquad \left.+ \lambda  \left[\hat{b}^\dagger_{n,\text{v}} L_{-,n} \hat{b}_{n+1,\text{p}} + \text{h.c.}\right]  \right\rbrace \nonumber\\ &\qquad +  \left( \text{other gauge-invariant terms} \right) \; .\nonumber 
\end{align}
With fermions instead of the bosons $\hat{b}$, the first line above is the Hamiltonian of lattice QED  with Wilson fermions. In this case, we can relate the constants $\chi, \Omega$ and $\lambda$ with the electric coupling $e$, the fermion mass $m$ and the lattice spacing $a$ as
\begin{align}
\chi \leftrightarrow \frac{ae^2}{2} \;, && \Omega \leftrightarrow m + \frac{1}{a} \;, && \lambda \leftrightarrow \frac{1}{a \sqrt{\ell\left(\ell + 1\right)}} \;,
\end{align}
where $\ell$ is the spin length of the gauge fields.

\section{The building block}
In the following, we focus on a single building block $n$ of the extended proposal, as realized in the present experiment. To this end, we consider only two matter field components, $ \hat{b}_\text{p} = \hat{b}_{L,n,0}$ and $ \hat{b}_\text{v} = \hat{b}_{L,n,1}$, together with a single spin $\hat{\mathbf{L}} = \hat{\mathbf{L}}_{n}$. The microscopic Hamiltonian conserves the total magnetization $\hat{M}$ and the total particle numbers $\hat{N}_L$, $\hat{N}_N$. To simplify the Hamiltonian of a single building block, we assume an initial state with fixed magnetization $M$ and particle numbers $N_N$, $N_L$. After some algebra, we obtain the building block Hamiltonian given in Eq.~(1) of the main text,

up to constants involving only the conserved numbers $N_N$, $N_L$ and $M$. The Hamiltonian is gauge-invariant because it commutes with the two reduced Gauß' law operators associated with neighbouring lattice sites,
\begin{align}
\hat{G}'_n = \hat{L}_{z} + \hat{b}_\text{p}^\dagger \hat{b}_\text{p} \;, && \hat{G}'_{n+1} = -\hat{L}_{z}  + \hat{b}_\text{v}^\dagger \hat{b}_\text{v} \; .
\end{align}
The parameters are given by
\begin{subequations}
	\begin{align}
	\Delta &=  -\left[\deltaL-\deltaN + \deltaLL - \deltaNN \right. \\ & \quad + \left. \frac{\chiNL}{2}(N_N-N_L) + 2M\left\lbrace -\chiLL-\frac{\chiNL}{2}\right\rbrace \right] \label{eqn82} \;,\\
	\chi &= -\chiNN-\chiLL-\chiNL \;,\\
	\lambda  & =  \XSCC \; ,
	\end{align}
\end{subequations}
where we abbreviated the effective interaction constants $\chiSS =  -\left[\XZeroZeroS +\XOneOneS-2\XOneZeroS\right]= \left[\XOneOneS - \XZeroZeroS\right]$, the mean-field energy shifts $\deltaSS =  \left[\XOneOneS-\XZeroZeroS\right] (N_s-1)=\chiSS(N_s-1)$ and the single-particle energy level differences $\deltaS = E_{s,1}-E_{s,0}$. The interaction constants are calculated from the relevant overlap integrals by rescaling the microscopic parameters as $X_{\alpha \beta}^s=I^s_{\alpha \beta}\frac{g^s_{\alpha\beta}}{2}$, $X_{\alpha \beta}^{Mis}=I^{Mix}_{\alpha \beta}g^{Mix}_{\alpha\beta}$ and $X^{SCC} = I^{Mix}_{10} g^{SCC}$ with
\begin{subequations}
	\begin{align}
	I^{s}_{\alpha \beta} &= \int_\mathbf{x} \Phi^2_{s,\alpha,n}(\mathbf{x}) \Phi^2_{s,\beta,n} (\mathbf{x}) \;, \\ I^{Mix}_{\alpha \beta} &= \int_\mathbf{x} \Phi^2_{N,\alpha,n}(\mathbf{x}) \Phi^2_{L,\beta,n} (\mathbf{x})\;,
	\end{align}
\end{subequations}
where we have chosen real basis functions $\Phi$.
From the known experimental parameters, we estimate the overlap integrals from the initial mean-field BEC wave-function $\Phi_{s,1,n}$ with all atoms of both species in the $m_F=1$ state by setting $\Phi_{s,1,n} = \Phi_{s,0,n}$. We have calculated the necessary wave-functions by imaginary-time propagation of the Gross-Pitaevskii equation corresponding to full microscopic Hamiltonian. In the experiment, we tune the magnetization $M$ through the initial value of the spin $L_z(0)$ and the magnetic field $B$. To quantify this dependence, we split $\Delta = \Delta_0 + \Delta_L L_{z,N}(0)/L_N + \Delta_B (B-B_\text{A})/B_\text{A}$, which holds for $(B-B_\text{A}) \ll B_\text{A}$ and fixed initial condition for the \text{Li} atoms.
With the given experimental and microscopic parameters, we obtain the following estimates:
\begin{subequations}
	\begin{align}
	\frac{\chi^\text{th}}{2\pi}  &\approx \SI{21.42}{\milli \hertz} \;,  \\ \frac{\lambda^\text{th}}{2\pi} &\approx  \SI{123.7}{\micro\Hz} \;, \\ \frac{\Delta^\text{th}_0}{2\pi}  &\approx -\SI{31}{\Hz} \;, \\\frac{\Delta^\text{th}_L}{2\pi}  &\approx  \SI{6.546}{\kilo\Hz} \;, \\ \frac{\Delta^\text{th}_B}{2\pi} &\approx -\SI{1.669}{\kilo\Hz}  \; .
	\end{align}
\end{subequations}

\section{Effective description of the building block dynamics}
We find that all experimental data is well described by the building block Hamiltonian in the mean-field approximation with a phenomenological damping term. The deviation from the mean-field building block can be understood from considering experimental imperfections, such as fluctuating initial conditions and spatial inhomogeneities, or quantum fluctuations. These lead to a decoherence of the observed oscillations and renormalize the building block parameters. The damping is characterized by a decoherence time scale $1/\gamma$, which is determined by fitting the envelope of the oscillation in Fig.~\ref{fig:gaugefielddynamics}B with an exponential decay, yielding $1/\gamma = 46(12)$ms, i.e. $\gamma / 2\pi = \SI{3.54(94)}{\hertz}$.

In practice, we solve the dynamics arising from the Hamilonian in Eq.~(1) of the main text in the mean-field approximation. For convenience, we rewrite it in terms of two coupled effective spins of length $L_{N}=N_N/2$ and $L_L=N_L/2$ by using the Schwinger representation also for Li.
Then the Hamiltonian becomes $H/\hbar= \chi L_{z,N}^2 + \Delta L_{z,L} + 2\lambda \left(L_{x,N}L_{x,L} + L_{y,N}L_{y,L} \right)$. We take the damping into account by modifying the resulting equations of motion as
\begin{subequations}
	\begin{align}
	\partial_t L_{x,N} &= - 2\chi L_{z,N}L_{y,N}  + 2\lambda L_{z,N} L_{y,L} - \frac{\gamma}{2} L_{x,N}\;, \\
	\partial_t L_{y,N} &= 2\chi L_{z,N}L_{x,N}  -2\lambda L_{z,N} L_{x,L} - \frac{\gamma}{2} L_{y,N}\;, \\
	\partial_tL_{z,N} &=  2\lambda (L_{y,N}L_{x,L} - L_{x,N}L_{y,L})\;, \\
	\partial_t L_{x,L} &=  -\Delta L_{y,L} + 2\lambda L_{z,L} L_{y,N} - \frac{\gamma}{2} L_{x,L}\;, \\
	\partial_t L_{y,L}&=  \Delta L_{x,L} - 2\lambda L_{z,L} L_{x,N} - \frac{\gamma}{2} L_{y,L} \;,\\
	\partial_t L_{z,L}&=  2\lambda (L_{y,L}L_{x,N} - L_{x,L}L_{y,N}) \;,
	\end{align}
\end{subequations}
with initial conditions $\mathbf{L}_{L}(0) = (0,0,-1) \times L_L$ and $\mathbf{L}_{N}(0) = (\cos \theta,0,\sin \theta)\times L_N$. Here, $\theta \in \left[ -\pi, \pi \right]$ is chosen in accordance to the quench that initiates the dynamics. We fit the numerical solution to the resonance data via the observable $N_\text{p}= L_{z,L} + L_L$. To this end, we assume that $\Delta = \Delta_0 + \Delta_L L_{z,N}(0)/L_N + \Delta_B (B-B_\text{A})/B_\text{A}$. This dependence of $\Delta$ on the initial spin $L_{z,N}(0)$ (for fixed $\mathbf{L}_{L}(0))$ and the magnetic field $B$ is analogous to the functional form derived in the previous section for an ideal building block. We obtain the effective model parameters $\chi, \lambda, \Delta_0$ and $\Delta_L$ from a single fit to the data shown in Fig.~\ref{fig:gaugefieldB}A with given $\gamma$. Fixing these parameters, $\Delta_B$ is determined by a second fit to the data of Fig.~\ref{fig:gaugefieldB}B-D.
Explicitly, the fits yield the following set of model parameters describing all experimental data:
\begin{subequations}
	\begin{align}
	\frac{	\chi }{2\pi} &= \SI{8.802(8)}{\milli\Hz} \;, \\ \frac{\lambda}{2\pi}  &=  \SI{16.4(6)}{\micro\Hz} \;, \\  \frac{\Delta_0 }{2\pi} &= -\SI{4.8(16)}{\Hz} \;, \\ \frac{\Delta_L }{2\pi} &=  \SI{2.681(1)}{\kilo\Hz} \;, \\ \frac{\Delta_B}{2\pi} &= -\SI{519.3(3)}{ \hertz}  \; .
	\end{align}
\end{subequations}

Comparing to the microscopic ab initio estimates gives
\begin{subequations}
	\begin{align}
	\chi^\text{th}/ \chi &\approx 2.4 \;, \\ \lambda^\text{th}/\lambda &\approx 7.6 \;, \\ \Delta_0^\text{th} / \Delta_0 &\approx 6.4 \;, \\ \Delta_L^\text{th} / \Delta_L &\approx 2.5 \;, \\ \Delta_B^\text{th} / \Delta_B &\approx 3.2 \; .
	\end{align}
\end{subequations}
The estimates are larger than the measured values by factors of about $2$ to $8$, which hints at a systematic uncertainty in the microscopic parameters that enter the ab initio calculation. Nevertheless, all values lie in the right order of magnitude. Importantly, the signs agree, such that the qualitative nature of the shift due to $B$ and $L_{z,N}$ is predicted in agreement with the experiment.

In terms of the two coupled spins, the resonance can also be understood more intuitively as follows. In our case, $L_N \approx 6 L_L$ and thus by angular momentum conservation the component $L_{z,N}$ can not change much (at most $\sim 17 \%$) in the course of the dynamics. Now, if the spins were not coupled, they would rotate in the $(x,y)$-plane with frequencies given by $2\chi L_{z,N}$ and $\Delta$. If the two frequencies were very different, the coupling $\propto \lambda$ would average out. Therefore $2\chi L_{z,N} \sim \Delta$ gives an approximate condition for the resonance (assuming approximately constant $L_{z,N}$). Solving this condition for $L_{z,N}$ with the fitted experimental paramters gives the following resonance estimates for the four magnetic fields employed in the experiment:
\begin{align}
\left.\frac{L_{z,N}}{L_N}\right|_\text{resonance} &= \frac{\Delta_0 + \Delta_B \left(B-B_\text{A}\right)/B_\text{A}}{\chi N_N- \Delta_L}\\ &\in   \left\lbrace  0.12(4), -0.18(4), -0.47(5), -0.78(6) \right\rbrace \;, \nonumber\\ B &\in \left\lbrace B_\text{A}, B_\text{B}, B_\text{C}, B_\text{D} \right\rbrace  \; . \nonumber
\end{align}
The actual peak position is not accurately reproduced because the employed resonance condition is not quantitative for the employed particle numbers and nonlinearities of the equations of motion modify the resonance. Nevertheless, the qualitative shift of the resonance peak is well captured.

\end{document}